\pdfoutput=1
\documentclass{JINST}
\usepackage[pdftex]{graphicx}
\usepackage{amsmath,amssymb,booktabs,url,verbatim}
\newcommand{\1}[1]{\, \mathrm{#1}} 

\title{Xenon Recirculation--Purification with a Heat Exchanger}

\author{K.~L.~Giboni$^a$\thanks{Corresponding author. E-mail: \texttt{kgiboni@astro.columbia.edu}}, E.~Aprile$^a$, B.~Choi$^a$, T.~Haruyama$^b$, R.~F.~Lang$^a$, K.~E.~Lim$^a$, A.~J.~Melgarejo$^a$, G.~Plante$^a$\\
\llap{$^a$}Columbia Astrophysics Laboratory, 550 W 120th St, New York, NY 10027, USA \\
\llap{$^b$}KEK, High Energy Accelerator Research Organization, 1-1 Oho, Tsukuba, Ibaraki 305-0801, Japan}

\abstract{Liquid-xenon based particle detectors have been dramatically growing in size during the last years, and are now exceeding the one-ton scale. The required high xenon purity is usually achieved by continuous recirculation of xenon gas through a high-temperature getter. This challenges the traditional way of cooling these large detectors, since in a thermally well insulated detector, most of the cooling power is spent to compensate losses from recirculation. The phase change during recondensing requires five times more cooling power than cooling the gas from ambient temperature to -100$^\mathrm{o}$C (173~K). Thus, to reduce the cooling power requirements for large detectors, we propose to use the heat from the purified incoming gas to evaporate the outgoing xenon gas, by means of a heat exchanger. Generally, a heat exchanger would appear to be only of very limited use, since evaporation and liquefaction occur at zero temperature difference. However, the use of a recirculation pump reduces the pressure of the extracted liquid, forces it to evaporate, and thus cools it down. We show that this temperature difference can be used for an efficient heat exchange process. We investigate the use of a commercial parallel plate heat exchanger with a small liquid xenon detector. Although we expected to be limited by the available cooling power to flow rates of about 2~SLPM, rates in excess of 12~SLPM can easily be sustained, limited only by the pump speed and the impedance of the flow loop. The heat exchanger operates with an efficiency of ($96.8\pm0.5$)\%. This opens the possibility for fast xenon gas recirculation in large-scale experiments, while minimizing thermal losses.}


\begin{document}

\section{Introduction}

Over the last 15~years, xenon detector technology made giant advances. Target masses are no more hundreds of grams, but are rapidly approaching a scale of several tons of liquefied xenon. Notably two improvements enabled this evolution: The application of the recirculation--purification scheme, and the development of powerful cryocoolers, specifically designed for liquid xenon temperatures. Both technologies were first developed for and realized in the MEG experiment~\cite{Adam:2009ci}, which uses a total of 2.7~tons of xenon.

In the recirculation--purification scheme, small amounts of liquid are extracted continuously from the detector, evaporated, and passed through a gas purifier. The clean gas is then recondensed into the detector. The purity of the liquid in the detector is thus controlled by two time constants. One stems from the outgassing of impurities from surfaces in contact with the liquid, and the other from the effectiveness of the purification filter. If recirculation is sufficiently fast, the purification outweighs the outgassing of impurities, and the liquid purity improves progressively, thus reaching substantially longer attenuation lengths for both scintillation light and ionization signals. This condition defines the minimal recirculation rate for experiments which run continuously for many months.

The outgassing rate increases as the detector size increases, since it becomes more difficult for large detectors to be baked under vacuum, and larger amounts of signal sensing devices and cabling are required in direct contact with the liquid. Therefore, purification of the liquid can be done efficiently only after filling the detector. To make this process more efficient, it is desirable to keep the latent heat from the evaporation of xenon within the system.

Pulse Tube Refrigerators~(PTRs), developed and optimized specifically for liquid xenon temperature, are a very convenient source of cooling power~\cite{Haruyama:2004xw,Haruyama:2004xx}. After their original design at KEK, they became commercially available from Iwatani Co. in a wide range of cooling powers, up to a present maximum of 200~W at~165K~\cite{Haruyama:2005vr}. For example, the Dark Matter search experiment XENON100, which contains more than 160~kg of liquid xenon, is cooled by one such cryocooler~\cite{Aprile:2010um}. Less than half of its cooling power is used to compensate heat leaking into the detector, whereas the other half remains available for recondensing xenon after purification. Since no heat exchanger is used in XENON100, the available cooling power allows a maximum recirculation rate of 10~standard liters per minute (SLPM), which corresponds to about 3~kg/hour of xenon.

For a future ton-scale detector, such as XENON1T with a total mass of about 2.5~tons, the time to pass the liquid through the purifier once at this rate would be over a month. Given that purification by recirculation is an exponential process, the timescale to get a reasonable purity would then be of the order of a year. Scaling the cooling system by a similar amount is however not practical. The above-mentioned PTR requires 6.5~kW of power for a helium compressor and another 5~kW for a water chiller. The cooling power of the PTR could be enhanced with a higher power compressor, but would soon saturate. The only solution would thus be an array of a large number of PTRs, which would obviously be very expensive and consume too much power, in particular for location in an underground laboratory.

The only viable solution appears to be the transfer of heat from the gas to the liquid in an efficient way, including the latent heat of the phase change. This can be achieved with commercially available heat exchangers as we demonstrate in the following.

\section{Balance of Cooling Power}\label{sec:balance}

The amount of heat to be transferred between the two (in/out) xenon streams can be calculated from the temperature difference, the heat capacitance, and the latent heat. Since we want to relate it to the flow rate, we express the heat in watt per standard liter per minute (W/SLPM). With $c_p\sim0.17\1{J\,g^{-1}\,K^{-1}}$ at 0.180~MPa and a xenon latent heat of 93~J/g~\cite{Lemmon2011:ni}, heating xenon gas from -97$^\mathrm{o}$C (176~K) to ambient temperature (293K) requires about 1.9 W/SLPM. However, evaporation alone requires about 9.0 W/SLPM and happens at the given boiling temperature. Hence, for the heat exchanger to be efficient, most of the heat has to be transferred at zero temperature difference. Otherwise, even a parallel plate heat exchanger, being the most efficient design, would save less than 20\% from heating the gas alone.

The occurrence of a phase transition is neither new nor unique. There are ways to overcome this problem, and one of them can be implemented naturally in the recirculation--purification scheme. The liquid is sucked through a thin tube from the detector by a gas recirculation pump, and then pressed through the rest of the purification system. Due to the underpressure on top of the liquid, the boiling point is lowered, and the liquid will evaporate, thus cooling the xenon on a line of constant enthalpy. This effect can provides the required temperature difference for efficient heat exchange, since the out-going stream of xenon is colder than the boiling point at normal detector pressure.

\section{Experimental Setup}

\subsection{Xenon Detector}

The tests reported here were performed with a setup designed for a detector that is optimized for a more precise measurement of the scintillation efficiency of low energy nuclear recoils in liquid xenon~\cite{Aprile:2008rc,Aprile:2006kx}. The detector vessel was used prior to instrumenting it with the time projection chamber and photomultiplier tubes. The total mass of xenon condensed in the vessel was about 1.5~kg. The detector vessel, including the cooling system and the heat exchanger, are all placed inside a vacuum chamber for thermal insulation. To further reduce the heat transfer by radiation, all parts are surrounded by standard superinsulation foil.

\subsection{Cooling}

\begin{figure}[!htbp]
\begin{center}
\includegraphics[width=.6\columnwidth]{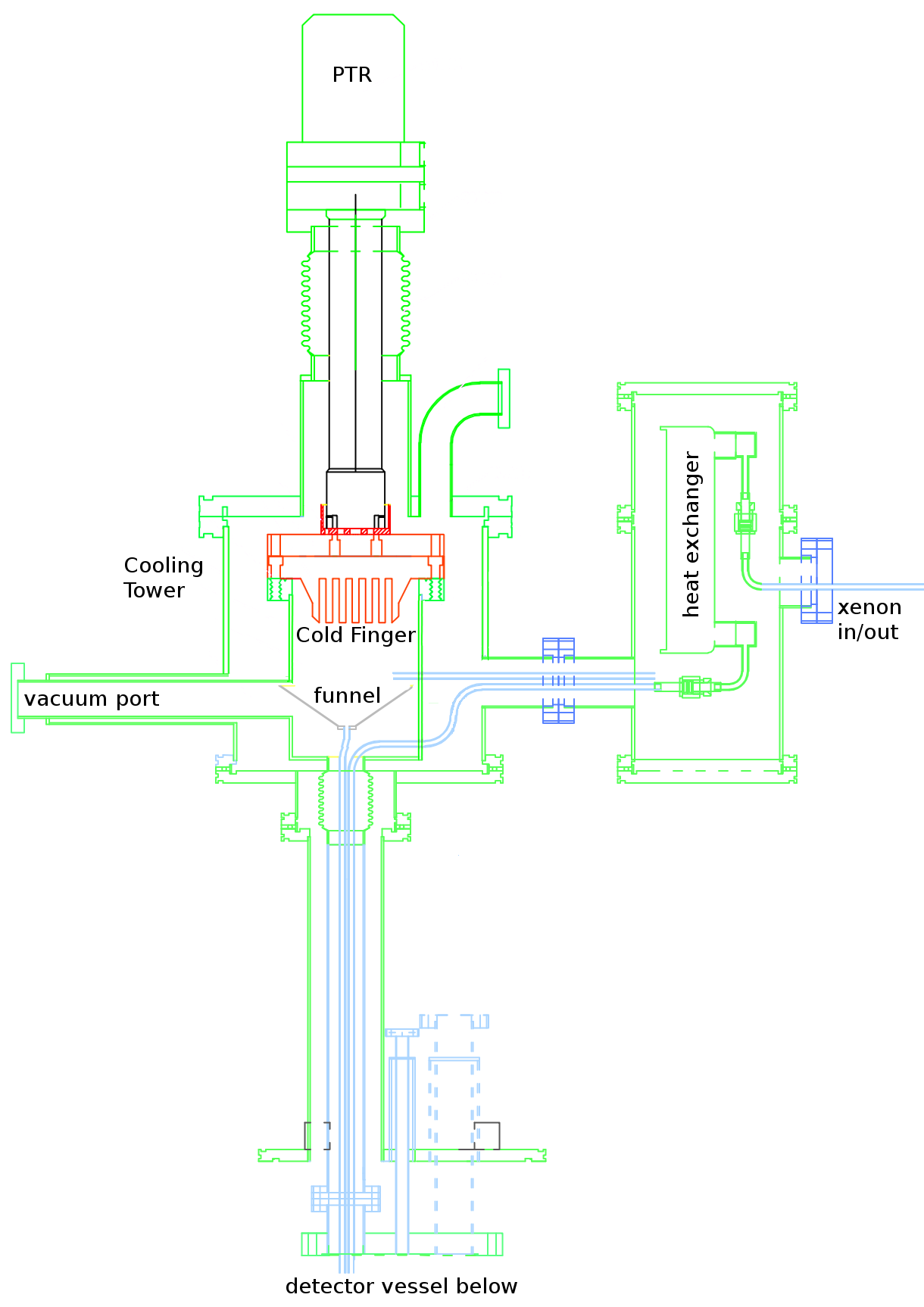}
\end{center}
\caption{Drawing of the Cooling Tower with PTR and funnel to collect condensed xenon. The heat exchanger is mounted in a separate vessel to the right. The detector vessel, not shown, is mounted below the Cooling Tower.}
\label{fig:cooling}\end{figure}

Figure~\ref{fig:cooling} shows a schematic of the cooling system, which follows the same principles already used in the XENON100 experiment. Xenon liquefaction does not take place in the detector vessel itself, but in a separate vessel, the so-called Cooling Tower, located about 50~cm above the detector volume.

The Cooling Tower hosts a PTR and is connected with the detector vessel by a double-walled tube, insulated by vacuum and superinsulation foil. Through this tube, xenon gas from above the detector volume can freely pass into the Cooling Tower. A copper Cold Finger connects the PTR to the detector volume and liquefies the xenon gas. An ohmic resistor, acting as a heater, is installed on the Cold Finger to compensate any excess cooling power. Pt100 sensors measure the temperature of the Cold Finger and provide a feedback to control the power that is supplied to the heater, so that the temperature of the system remains constant. The Cold Finger is sealed at its perimeter to the inner walls of the Cooling Tower, and protrudes inwards with a fin-like structure for better cold transfer to the xenon gas. The cold head of the PTR itself is thus within the insulation vacuum. This mounting detail is important to keep heater, temperature sensors, and wiring out of the xenon volume for purity reasons. In addition, it allows access to the PTR without exposing the inner detector vessel to air. Xenon condenses on the Cold finger, is collected by a funnel, and freely runs through a small tube to finally drip into the detector.

An Iwatani PDC08 PTR with a 700~W air-cooled helium compressor was used in these tests. To measure the available cooling power, the chamber was evacuated and the heater was used to control the temperature of the Cold Finger to 173~K. The required heater power to achieve this equilibrium was 29~W and corresponds to the available cooling power of the PTR. The slightly higher cooling power compared to a similar system at KEK is explained by the 60~Hz operation in our laboratory in the US. At the operating temperature, the xenon vapor pressure in the detector was about 0.18~MPa absolute. Without recirculation, the heater on the Cold Finger required 16~W to keep the liquid xenon at a constant temperature. Thus, 13~W of cooling power were used to counteract residual heat transfer through imperfect thermal insulation.

\subsection{Recirculation}

\begin{figure}[!htbp]
\begin{center}
\includegraphics[width=1.0\columnwidth]{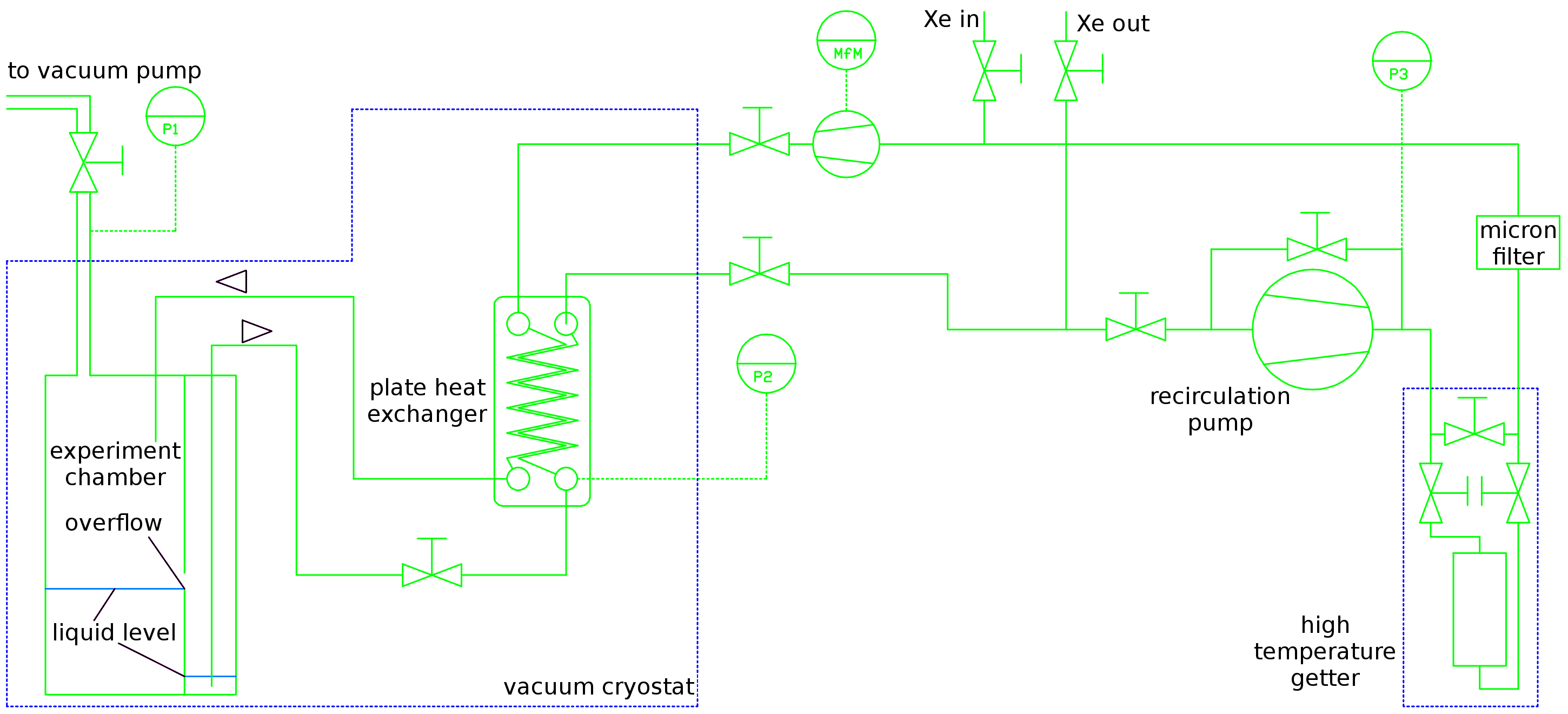}
\caption{Schematic drawing of the recirculation-purification system.}
\label{fig:gassystem}
\end{center}
\end{figure}

The recirculation system consists of a diaphragm pump~\cite{equipment:mx-808st-s}, a high temperature getter~\cite{equipment:ps4-mt3-r1} to purify the xenon, a micron filter to remove dust from the gas flow, and a flow meter~\cite{equipment:hfc-302}. After passing through these items in sequence, the xenon gas is transferred back to the detector. Figure~\ref{fig:gassystem} shows a schematic of the recirculation system.

Without a heat exchanger, the flow rate is limited by the available cooling power, and about 2~SLPM can routinely be achieved. The elements of the recirculation loop impose weaker constrains in the flow. The 1/4" tubing is no significant limitation in our case, since the pump can provide a sufficient pressure difference of 0.2~MPa to overcome the flow resistance. The pump itself has a maximum through-put of 28~SLPM. The getter is specified with 5~SLPM for correct operation, although higher flow rates are possible and were used during the present tests at the expense of reduced purification. Finally, the flow meter is calibrated up to 10~SLPM by the supplier, and does not display any value above 19~SLPM. The most stringent limitation on the achievable flow rate is however caused by a micron filter within the getter module. These limitations can be overcome in future large systems if these are equipped with larger tubing, stronger pump, faster flow meter, and higher capacity getter, all readily available components.

\subsection{Heat Exchanger Module}

In previous xenon systems with recirculation--purification, heat had to be introduced to the system to evaporate xenon and warm it to room temperature, but at the same time, a similar amount of heat had to be extracted from the purified gas for cooling and liquefaction. Considering the complexity and the electrical power required by a large cooling system, this is only practical for small systems. In order to test an alternative solution, we coupled the incoming and outgoing xenon lines with a commercial heat exchanger.

This heat exchanger is a flat plate heat exchanger~\cite{equipment:fg3x8-20}, mounted for simplicity between two G-10 plates in a separate vacuum vessel, right next to the vacuum vessel of the Cooling Tower. It has 20 stainless steel plates, copper-brazed together, to form the exchanger structure. Cooling power losses through the G-10 plates are estimated to be negligible. Furthermore, the heat exchanger is quite compact and light weight and could be held only by the connecting pipes, thus avoiding additional heat influx to the heat exchanger. The all-metal construction is expected to be sufficiently clean in order to fulfill the stringent purity requirements of a liquid xenon detector. Baking at up to 450$^\mathrm{o}$C (723~K) is possible, although not attempted for the tests reported here.

The heat exchanger model we used has large 3/4" NPT openings. The change in cross section from our standard 1/4" tubing also causes the flowing xenon to expand and cool. Since it was not obvious if this effect were sufficient to create the required pressure gradient, a valve was mounted within the insulation vacuum space, to form a variable orifice valve. This valve, and also another valve in front of the pump, could also be used to control and adjust the xenon flow. After the orifice valve, a pressure sensor was used to measure the pressure drop.

\section{Results and Discussion}

The detector was operated at different flow rates, and the power supplied to the heater was measured. High flow rates were easily achieved, exceeding by far the 1.5~SLPM that were expected from the 16~W of unused cooling power of the PTR. When throttling the flow below 4~SLPM, the pressure gauge indicated a pressure drop of more than 0.1~MPa across the orifice valve. At such pressure differences, the xenon might cool down too much and freeze, blocking the thin 1/4" tube. Thus, at these low flow rates, a valve in front of the pump was used to reduce the gas flow. Since this valve comes after the heat exchanger, the xenon gas is already at ambient temperature.

At the other end of the probed flow range, the orifice valve could be opened completely without any apparent change in heat exchanger efficiency, although the pressure drop across the valve was negligible. The explanation is the large mismatch in tube cross section by more than a factor~20 at the input of the heat exchanger: the orifice valve was not needed at all. As a cross check, we verified that the efficiency of the heat exchanging process does not depend on which of the two available valves limits the flow.

\begin{figure}[!htbp]
\begin{center}
\includegraphics[width=0.6\columnwidth,clip,trim=90 150 90 220]{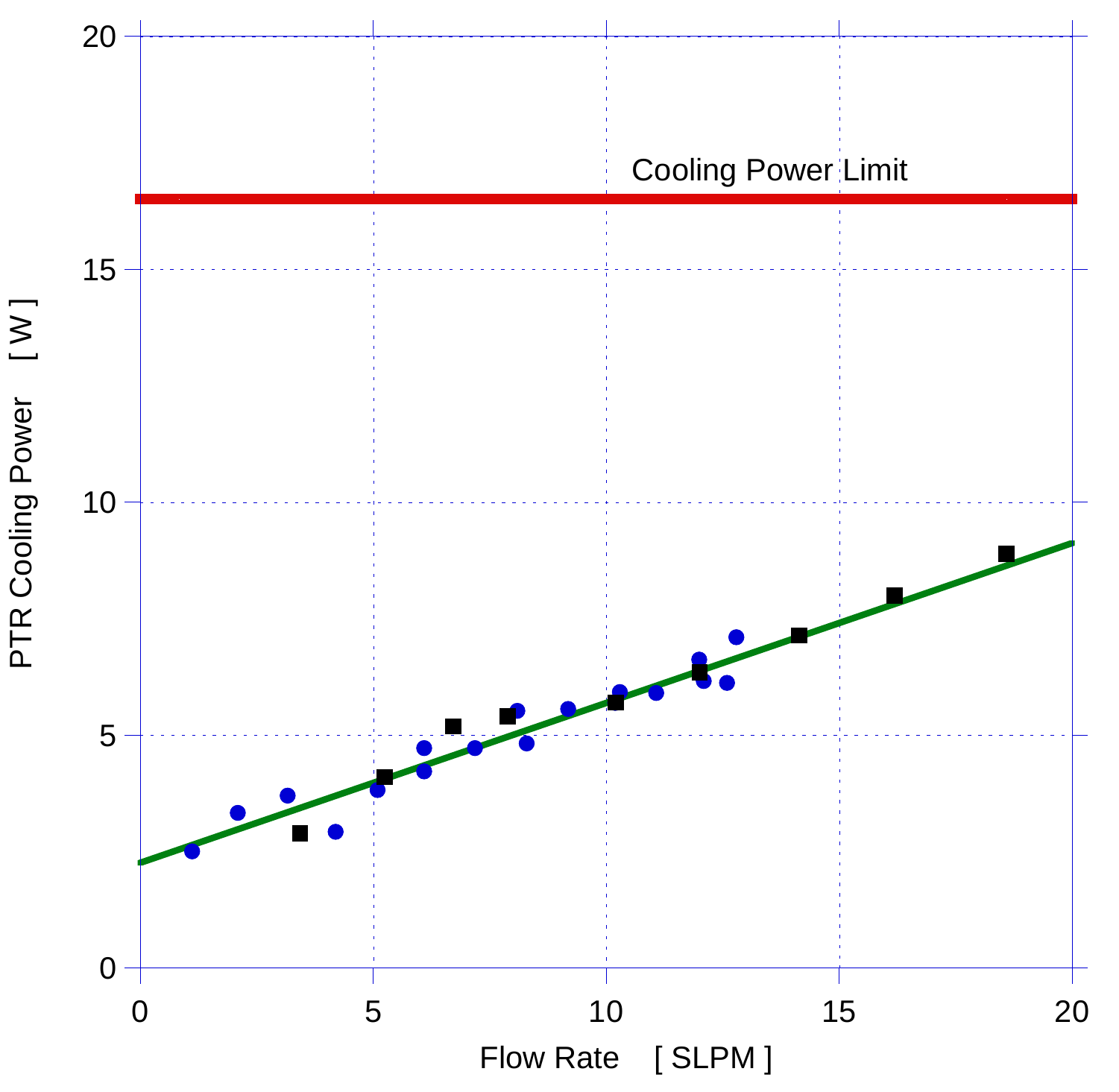}
\caption{Cooling Power $P$ versus Flow Rate $r$. Squares (black points) are for the recirculation of gas evaporated from the liquid, circles (blue points) are from a run where only gas above the liquid was recirculated. The horizontal (red) line indicates 16~W, whereas the inclined line (green) is a linear fit $P=2.26\1{W}+0.34 \1{W/SLPM} \:r$ to the squared (black) data points.}
\label{fig:data}
\end{center}
\end{figure}

Figure~\ref{fig:data} shows the required cooling power $P$ versus flow rate $r$. The cooling power is computed as PTR heating power minus the 'no flow' power value of 16~W. A sudden change in flow rate would cause the heating power to react only gradually. Also, since the flow rate was only measured, but not controlled, both the cooling power and the flow rate would drift towards equilibrium very slowly. In the present system, a reliable pair of measurement points would require at least 2~hours before a new equilibrium was achieved. Thus, the system was allowed to stabilize for at least 6~hours, often longer or over night, before recording the values shown in figure~\ref{fig:data}. The residual error is smaller than the spread of the points from the line, which can be mostly attributed to temperature changes in the laboratory. Data for this study was taken as a side activity over a very long time, and would have been impossible without the high degree of stability of the PTR cooling.

The measured data points (circles) fit a straight line, from the minimum of about 1~SLPM to a maximum of nearly 13~SLPM. We interpret the slope as a measure of the efficiency of the heat transfer process. Comparing the slope of 0.34~W/SLPM to the 10.6~W/SLPM that are needed to cool and liquefy 1~SLPM of xenon from room temperature, we can then conclude that the efficiency of the heat exchange is ($96.8\pm0.5$)\%. The offset of the data is the cooling power required with no recirculation. This cooling power has to compensate for all the thermal losses in the connecting lines that are present despite the insulation vacuum and wrapping with superinsulation foil. The losses of the heat exchanger and its mounting are included in this offset, but are estimated to be small.

In a subsequent test, the chamber was filled with less liquid xenon, so that only gas on top of the liquid was recirculated. The results are shown as square points in figure~\ref{fig:data}. These points fit the same line determined in the previous measurement. The gas has about the same temperature as during liquid circulation, i.e. the heat losses of the system are very similar and we would expect the same offset. The sudden widening of the pipe cross section cools the gas by expanding it, and the temperature difference achieved in this way is sufficient to operate the heat exchanger with the same efficiency.

We conclude that this efficiency must be the maximum efficiency of this particular heat exchanger model. Increasing the flow rate will reduce the efficiency at some point, but these limitations lie outside of the flow range accessible in the present tests. By-passing the getter, but not the micron filter, did increase the maximum flow rate from about 13~SLPM to 18~SLPM. When both the getter and the micron filter were bypassed, the display of the flow meter limited the measurements to a maximum of 19~SLPM. The expected maximum recirculation rate that we calculate from the available cooling power is 35~SLPM, even higher than the maximum throughput of the pump of 28~SLPM.

We note that the tube coming out from the detector, after the heat exchanger, is free of the usual accumulation of ice or even water condensation. The gas exits the detector with the same temperature with which it re-enters, close to the ambient temperature in the lab.

\section{Conclusions}

The common procedure of using gaseous recirculation--purification for a liquid xenon detector appears ideal for the use of a heat exchanger: when evaporating the liquid from the detector, a large amount of heat has to be supplied to the xenon, whereas when the gas is returned after purification the same amount of heat has to be extracted from the gas. In addition, input and output flow rates are equal, since the gas is contained in a closed loop. However, 90\% of the required heat is latent heat for the phase change, and the exchange occurs at zero temperature difference. On the other hand, if the xenon is evaporated at a lower pressure than recondensation, then evaporation cools the xenon, and the phase change takes place with a sufficient temperature difference to recover the heat with a very high efficiency through a commercial heat exchanger. Since a pump is present for continuous gas purification anyway, the prerequisites for the use of a heat exchanger can easily be fulfilled.

In a small test system, with limited cooling power, we successfully demonstrated the use of a commercial heat exchanger to recover a significant fraction of cooling power. The efficiency of the system is measured to be ($96.8\pm0.5$)\%. The addition of the heat exchanger did not significantly add to the complexity of the system, nor to the cost of the gas handling system. Although for the present tests the detector was not yet instrumented, we have subsequently operated the instrumented detector with the cooling and purification system described here, with a high degree of reliability over extended periods of time. In addition, all specifications of the heat exchanger, such as the allowed temperature range or maximum pressure rating, exceed by far the values commonly encountered in a liquid xenon system.

In future large xenon systems, the recirculation speed cannot be chosen freely. For a given total xenon mass the characteristic time of recirculation is how often all the volume can be moved through the purification system once. This time constant has to be compared with the outgassing rate of the detector. For large detectors, instrumented with many photomultipliers immersed in the liquid volume, it will be impractical or impossible to bake and thoroughly outgas all construction elements. If the outgassing is faster than the purification, impurities will accumulate. Thus, a minimum recirculation speed needs to be sustained. For these detectors, a heat exchanger will become a necessity to allow continuous gas purification.

\section*{Acknowledgements}

We gratefully acknowledge support from the National Science Foundation through grants PHY-0705337 and PHY-0904220.


\begin{thebibliography}{10}

\bibitem{Adam:2009ci}
J.~Adam {\em et~al.}, {\it {A limit for the mu -> e gamma decay from the MEG
  experiment}},  {\em Nucl. Phys.} {\bf B834} (2010) 1--12,
  [\href{http://arxiv.org/abs/0908.2594}{{\tt arXiv:0908.2594}}].

\bibitem{Haruyama:2004xw}
T.~Haruyama, K.~Kasami, H.~Inoue, S.~Mihara, and Y.~Matsubara, {\it Development
  of a high-power coaxial pulse tube refrigerator for a liquid xenon
  calorimeter},  {\em AIP Conference Proceedings} {\bf 710} (2004), no.~1
  1459--1466.

\bibitem{Haruyama:2004xx}
T.~Haruyama {\em et~al.}, {\it {High-power pulse tube cryocooler for liquid
  xenon particle detectors}},  {\em Cryocoolers} {\bf 13} (2004) 689--694.

\bibitem{Haruyama:2005vr}
T.~Haruyama {\em et~al.}, {\it {LN-2 free operation of the MEG liquid xenon
  calorimeter by using a high-power pulse tube cryocooler}},  {\em Advances in
  Cryogenic Engineering 51, AIP Conf. Proc.} {\bf 823} (2006) 1695--1702.

\bibitem{Aprile:2010um}
E.~Aprile {\em et~al.}, {\it {First Dark Matter Results from the XENON100
  Experiment}},  {\em Phys. Rev. Lett.} {\bf 105} (2010) 131302,
  [\href{http://arxiv.org/abs/1005.0380}{{\tt arXiv:1005.0380}}].

\bibitem{Lemmon2011:ni}
E.~Lemmon, M.~McLinden, and D.~Friend, {\em NIST Chemistry WebBook, NIST
  Standard Reference Database Number 69}, ch.~Thermophysical Properties of
  Fluid Systems.
\newblock National Institute of Standards and Technology, Gaithersburg MD,
  2011.
\newblock {\url{http://webbook.nist.gov/chemistry/fluid/}}.

\bibitem{Aprile:2008rc}
E.~Aprile {\em et~al.}, {\it {New Measurement of the Relative Scintillation
  Efficiency of Xenon Nuclear Recoils Below 10 keV}},  {\em Phys. Rev.} {\bf
  C79} (2009) 045807, [\href{http://arxiv.org/abs/0810.0274}{{\tt
  arXiv:0810.0274}}].

\bibitem{Aprile:2006kx}
E.~Aprile {\em et~al.}, {\it {Simultaneous Measurement of Ionization and
  Scintillation from Nuclear Recoils in Liquid Xenon as Target for a Dark
  Matter Experiment}},  {\em Phys. Rev. Lett.} {\bf 97} (2006) 081302,
  [\href{http://arxiv.org/abs/astro-ph/0601552}{{\tt astro-ph/0601552}}].

\bibitem{equipment:mx-808st-s}
Model MX-808ST-S, Enomoto Micro Pump Mfg. Co., Ltd., Ebisunishi Shibuya-Ku,
  Tokyo, Japan, \url{http://www.emp.co.jp}.

\bibitem{equipment:ps4-mt3-r1}
Model MonoTorr PS4-MT3-R1, SAES Pure Gas, Inc., San Luis Obispo, California,
  USA, \url{http://www.saesgetters.com}.

\bibitem{equipment:hfc-302}
Model HFC-302, Teledyne Hastings Instruments, Hampton, Virginia, USA,
  \url{http://www.teledyne-hi.com}.

\bibitem{equipment:fg3x8-20}
Model FG3X8-20, GEA PHE Systems NA, Inc., York, Pennsylvania, USA,
  \url{http://www.gea-phe.com}.

\end{thebibliography}

\providecommand{\href}[2]{#2}\begingroup\raggedright\endgroup

\end{document}